\documentclass[preprintnumbers,prd,onecolumn,showpacs,floatfix,preprintnumbers,letterpaper,amsmath,amssymb,nofootinbib,superscriptaddress]{revtex4}
\usepackage{graphicx}
\usepackage{epsfig}
\usepackage{bm}
\usepackage{hyperref}
\usepackage{color}
\usepackage{amsmath}
\usepackage{amssymb}
\usepackage{amsthm}
\usepackage{amsfonts}
\usepackage{latexsym,float,url}

\def\be{\begin{equation}}
\def\ee{\end{equation}}
\def\ba{\begin{eqnarray}}
\def\ea{\end{eqnarray}}
\renewcommand{\(}{\left(}
\renewcommand{\)}{\right)}
\renewcommand{\[}{\left[}
\renewcommand{\]}{\right]}
\renewcommand{\ee}{\tilde e}
\renewcommand{\tt}{\tilde T}
\renewcommand{\SS}{\tilde S}

\begin{document}

\title{Acausality and Nonunique Evolution in Generalized Teleparallel Gravity}

\author{Keisuke Izumi}
 \email{izumi@phys.ntu.edu.tw}
 \affiliation{Leung Center for Cosmology and Particle Astrophysics,
 National Taiwan University, Taipei 10617, Taiwan}

\author{Je-An Gu}
 \email{jagu@ntu.edu.tw}
 \affiliation{Leung Center for Cosmology and Particle Astrophysics,
 National Taiwan University, Taipei 10617, Taiwan}

 \author{Yen Chin Ong}
\email{ongyenchin@member.ams.org}
 \affiliation{Leung Center for Cosmology and Particle Astrophysics,
 National Taiwan University, Taipei 10617, Taiwan}
 \affiliation{Graduate Institute of Astrophysics, National Taiwan
University, Taipei 10617, Taiwan}

\begin{abstract}
We show the existence of \emph{physical} superluminal modes and acausality in the Brans-Dicke type of extension of teleparallel gravity that includes $F(T)$ gravity and teleparallel dark energy as special cases. We derive the characteristic hypersurface for the extra degrees of freedom in the theory, thereby showing that the time evolution is not unique and closed causal curves can appear. Furthermore, we present a concrete disastrous solution in Bianchi type I spacetime, in which the anisotropy in expansion can be any function of time, and thus anisotropy can emerge suddenly, a simple demonstration that the theory is physically problematic.
\end{abstract}
\maketitle

\section{Introduction}\label{0}

Teleparallelism (or ``Fernparallelismus'') was proposed by Einstein~\cite{Einstein,Sauer} to unify gravitation and electromagnetism.
Unlike general relativity (GR), which employs the Levi-Civita connection, a spacetime in teleparallelism is equipped with the Weitzenb\"{o}ck connection~\cite{Weitzenbock, Weitzenbock2}, which gives zero curvature but nonvanishing torsion.
Due to the zero-curvature of the connection, the parallel transport of a vector is independent of path.
Indeed, \textit{teleparallel} means ``parallel at a distance.''
Despite Einstein's earlier failure to formulate the unified theory,
the idea of teleparallelism survived as a pure theory of gravity.
It turns out that Einstein's GR can be reformulated with teleparallelism, called teleparallel equivalent of general relativity (TEGR)~\cite{Baez,Hayashi,Pereira,Kleinert,Sonester}.
Despite such equivalence, teleparallelism gives a different perspective: Teleparallel gravity can be regarded as a gauge theory of gravity; in addition, it has an advantage in describing gravitational (quasilocal) energy~\cite{Nester1989,Blagojevic:2002du,Blagojevic:2000pi,Hehl:2012pi,Hehl:1994ue,Itin:1999wi,Cho:1975dh,Moller:1961jj,deAndrade:2000kr}.

To explain the observed cosmic acceleration of the present universe~\cite{Perlmutter:1998np,Riess:1998cb}, a variety of modified theories of gravity have been proposed, and TEGR has also been generalized in several ways analogous to the generalizations of GR. Although TEGR and GR are equivalent, their generalizations are not. For example, $F(T)$ gravity~\cite{0,1} replaces in the action the linear function of the torsion scalar $T$ with a general function, analogous to the approach of $F(R)$ gravity, while the Brans-Dicke type of extension introduces a scalar field $\phi$ nonminimally coupled to the torsion scalar, analogous to the scalar-tensor theory \cite{BD0}. Such Brans-Dicke type theory dubbed ``teleparallel dark energy'' has been proposed \cite{Geng:2011aj,Geng:2011ka}, and the Friedmann-Lema\^itre-Robertson-Walker (FLRW) cosmology in this theory widely studied (e.g., see \cite{Gu:2012ww,Geng:2013uga} and references therein).

Like all other new theories, it is crucial to check whether these generalized theories of teleparallel gravity are healthy. That is, we should investigate the behavior of the physical degrees of freedom (DoFs) therein and check if the theory is free of ghost, tachyonic behavior, and instabilities, among other potential problems.
(Note that ``degrees of freedom'' referred in the present paper are not necessarily physical unless specified explicitly.)
A generic theory of teleparallel gravity constructed of tetrad vectors has more DoFs than a metric theory like GR.
In TEGR the extra DoFs turn out to be inactive and do not change the physics of GR, mainly due to the local Lorentz symmetry. 
Nevertheless, in generalized theories of teleparallel gravity the gravity action is generically noninvariant under local Lorentz transformation~\cite{Li:2011wu}.
This lack of symmetry may activate the extra DoFs, making them dynamical, physical, and even out of control.
The Dirac constraint analysis by Li et al.~\cite{Li} showed that $F(T)$ gravity has three extra DoFs.
In a careful second-order linear analysis, the extra DoFs were shown to disappear on the FLRW background~\cite{IzumiOng}.\footnote{The disappearance of the extra DoFs on the Minkowski background has been pointed out in \cite{Li}.} One therefore needs to go beyond linear analysis to examine the behavior of the extra DoFs.

Nonlinear analysis of $F(T)$ gravity has been performed in \cite{Ong:2013qja}. This work established the existence of superluminal (in fact, \emph{infinite speed}) propagation by using the method of characteristics~\cite{CH}, a solid method widely used to analyze causal structure in gravity theories including GR. 
Moreover, specifically in a FLRW universe, the authors have shown the nonuniqueness of time evolution from the initial conditions prescribed on a constant-time hypersurface.
In particular, they presented an \emph{exact solution} of $F(T)$ gravity where one of the DoFs can be any function of time: 
\begin{eqnarray}\label{theta1}
&&e^0_{~\mu} dx^{\mu}= \cosh \theta (t) dt + a(t) \sinh \theta (t) dx,\nonumber\\
&&e^1_{~\mu} dx^{\mu}= \sinh \theta (t) dt + a(t) \cosh \theta (t) dx, \nonumber\\
&&e^2_{~\mu} dx^{\mu}= a(t) dy, \label{ansatz}\\
&&e^3_{~\mu} dx^{\mu}= a(t) dz, \nonumber
\end{eqnarray}
where $a(t)$ is the scale factor of the FLRW metric and $\theta(t)$ the extra DoF which can become out of control. For example, we can freely choose the form of the function $\theta (t)$ such that $\theta (t)=0$ for $t<0$ and $\theta (t) \neq 0$ for $t>0$,
thereby having a universe where torsion can suddenly emerge seemingly from nothingness on the spacelike hypersurface at time $t=0$.

Although it has been explained in \cite{Ong:2013qja} that the extra DoFs with such illness are \emph{not gauge} DoFs and therefore potentially physical, it is not made explicit whether this illness truly infects the physical world, i.e., affecting the physical quantities such as the Hubble expansion rate. To fully establish the illness of the theory, we need to check whether the extra DoFs are truly physical, i.e., coupled to the known physical fields such as metric and those in the Standard Model of particle physics, either directly or indirectly. If ill-behaved DoFs are completely decoupled from the physical ones, they simply dwell in their own world and make no harm to our physical world.

The extra DoFs in teleparallel gravity correspond to the freedom to have different tetrads which give the same metric.
In a generic teleparallel theory there can be more than one tetrad taht solves the field equations, and due to the lack of local Lorentz symmetry, these tetrads are considered distinct if they are not related by a \emph{global} Lorentz transformation.
That is, a generic teleparallel theory is a theory of preferred frame. A good teleparallel theory should \emph{not} admit more than one such global frame that parallelizes the spacetime. From this viewpoint the explicit solution given above casts doubt on the validity of the teleparallel theory. However, it is not clear how to experimentally distinguish the tetrads that correspond to the \emph{same} metric. Thus one reasonably wonders if the acausal\footnote{Acausality is a stronger statement than superluminality; we  explain why there is indeed acausality in the theory in the discussion.} DoFs are actually decoupled from the physical DoFs and therefore will neither change physics nor violate causality of the physical world.

This motivates the present work. We show that the problem is indeed physical, i.e., the ill-behaved extra DoF does couple to the metric DoFs and therefore is clearly physical. Beyond $F(T)$ gravity, we extend our analysis to a broader class of teleparallel theory, namely the Brans-Dicke theory that includes $F(T)$ gravity and teleparallel dark energy as special cases.

The organization of this paper is as follows.
In Sec.~\ref{2} we review the teleparallel formulation of gravity and TEGR. We then introduce two generalized theories of teleparallel gravity, namely $F(T)$ gravity and the Brans-Dicke type of extension, and show that the former is a special case of the latter.
In Sec.~\ref{4}, we analyze the equation of motion (EoM) for a general tetrad in the Brans-Dicke theory of teleparallel gravity, and derive the characteristic hypersurface for the extra DoFs.
We find that the constant-$\phi$ hypersurfaces are always characteristic hypersurfaces. 
In Sec.~\ref{5} we consider a simple case with the Bianchi type I metric for the demonstration of the acausality in teleparallel gravity. We present the nonuniqueness of time evolution in the gauge-invariant metric components, thereby showing that the acausal modes are truly physical.
Finally, we summarize with discussion in Sec.~\ref{6}.

We use the following notation for indices.
The Greek letters $\{\mu,\nu,\cdots\}$ are the indices for four-dimensional spacetime [i.e.\ $(t,x,y,z)$], while $\{i,j,\cdots\}$ are the indices for three-dimensional space [i.e.\ $(x,y,z)$].
For $(x,y)$ direction, we use $\{p,q,\cdots\}$.
The Latin letters $\{A,B,\cdots\}$ label tetrad vectors and run from $0$ to $3$, while
$\{I,J,\cdots\}$ run from $1$ to $3$. We use the  signature convention $(-,+,+,+)$.

\section{Teleparallel Gravity} \label{2}

In teleparallel gravity, the fundamental dynamical variables of
gravity are tetrad vectors $e^A{}_\mu$. The metric and the
metric-compatible Weitzenb\"{o}ck connection invoked therein
are constructed from the tetrad vectors as
\begin{eqnarray}
g_{\mu\nu} &=& \eta_{AB}e^A{}_\mu e^B{}_\nu \, , \label{metric} \\
\Gamma^\lambda{}_{\mu\nu} &=&
e_A{}^\lambda\partial_\nu e^A{}_\mu
= -e^A{}_\mu\partial_\nu e_A{}^\lambda \, . \label{connection}
\end{eqnarray}
The Weitzenb\"{o}ck connection always gives zero curvature but
can give nonzero torsion. The antisymmetric part of the
connection gives the torsion tensor, from which the gravitational action will be constructed:
\begin{eqnarray}
T^\lambda{}_{\mu\nu}:= \Gamma^\lambda{}_{\nu\mu}-\Gamma^\lambda{}_{\mu\nu}
=e_A{}^\lambda\partial_\mu e^A{}_\nu-e_A{}^\lambda\partial_\nu e^A{}_\mu \, .
\label{Tdef}
\end{eqnarray}
The difference between the Weitzenb\"{o}ck connection and the
Levi-Civita connection is given by the contortion tensor
\begin{eqnarray}
K^{\mu\nu}{}_\rho :=-\frac{1}{2}\(
T^{\mu\nu}{}_\rho-T^{\nu\mu}{}_\rho-T_\rho{}^{\mu\nu}
\).
\end{eqnarray}

One can construct the teleparallel gravity equivalent to
GR~\cite{Baez,Hayashi,Pereira,Kleinert,Sonester} with the
gravity action (in the unit $c=1$)
\begin{eqnarray}
S=-\frac{1}{2\kappa} \int d^4x~ |e| T,  ~~\kappa = 8\pi G,
\label{TEGR}
\end{eqnarray}
where $e=\det \left(e^A{}_\mu\right)$ and the torsion scalar $T$
is constructed from the torsion and contortion tensors as
\begin{eqnarray}
T &:=& S_\rho{}^{\mu\nu}T^\rho{}_{\mu\nu} \, , \\
\textrm{where} \quad
S_\rho{}^{\mu\nu} &:=& \frac{1}{2}\(
K^{\mu\nu}{}_\rho+\delta^\mu_\rho
T^{\lambda\nu}{}_\lambda-\delta^\nu_\rho T^{\lambda\mu}{}_\lambda
\).\label{Sdef}
\end{eqnarray}

Motivated by the discovery of the accelerated expansion of the present universe, recently TEGR has been generalized in the same spirit as the generalizations of GR.
In the present paper we consider two generalized theories of teleparallel
gravity, namely $F(T)$ gravity and the Brans-Dicke type of extension, with the following actions respectively:
\begin{eqnarray}
S_{FT} &=& -\frac{1}{2\kappa} \int d^4x ~|e| F(T),\label{F}\\
S_{BD} &=& -\frac{1}{2\kappa} \int d^4x ~|e|\[ f(\phi) T
-\alpha \( \partial_\mu \phi \)\(\partial^\mu \phi\) -V(\phi)\], \label{BD}
\end{eqnarray}
where $F(T)$ is an arbitrary function of the torsion scalar,
$f(\phi)$ and $V(\phi)$ arbitrary functions of the scalar field
$\phi$, and $\alpha$ a constant.
Despite the same spirit, these generalizations of TEGR give
different dynamics from their modified GR counterparts. The difference in dynamics
stems from the difference between the torsion scalar and the
Ricci scalar:
\begin{eqnarray}
\overset{\scriptscriptstyle \; L}R =
-T-2\overset{\scriptscriptstyle L\;\; }{\nabla^\mu} T^\nu{}_{\mu\nu} \, ,
\label{R=T}
\end{eqnarray}
where $\overset{\scriptscriptstyle \; L}R$ and
$\overset{\scriptscriptstyle L\;\; }{\nabla^\mu}$ are,
respectively, the Ricci scalar and the covariant derivative corresponding to the Levi-Civita connection. The difference is a divergence term that gives no effect on the dynamics in GR and TEGR.
However, this divergence term does make difference in the generalized theories.

We claim that the teleparallel theory with action $S_{FT}$ is actually a special case of $S_{BD}$.
The action $S_{FT}$ can be
recast into a different form by introducing auxiliary fields
$\psi$ and $\phi$:
\begin{eqnarray}
S_{FT}=-\frac{1}{2\kappa} \int d^4x ~|e| \[F(\phi)-\psi(\phi-T)\]. \label{F-}
\end{eqnarray}
The EoM from the variation with respect to $\psi$ requires
$\phi = T$, which, upon substitution into the action, reduces the
recast action back to the original one in Eq.~(\ref{F}).
In addition, the variation with respect to $\phi$ gives a
constraint equation:
\begin{eqnarray}
\psi = \frac{dF}{d\phi},
\end{eqnarray}
with which the action in Eq.~(\ref{F-}) can be rewritten as
\begin{eqnarray}
S_{FT}=-\frac{1}{2\kappa} \int d^4x~ |e|\[ \frac{dF}{d\phi} T
-\(\phi\frac{dF}{d\phi}-F \)\]. \label{14}
\end{eqnarray}
This action is explicitly a special case of $S_{BD}$, with
$\alpha=0$, $f(\phi)=dF/d\phi$ and $V(\phi)=\phi (dF/d\phi)-F$.
In addition, $S_{BD}$ clearly includes teleparallel dark energy~\cite{Geng:2011aj,Geng:2011ka}, in which the action is
\begin{eqnarray}
S_{TDE}=- \int d^4x ~|e|\[\frac{T}{2\kappa} + \frac{1}{2}\xi T \phi^2 - \frac{1}{2}\( \partial_\mu \phi \)\(\partial^\mu \phi\) -V(\phi)\].
\end{eqnarray}
Therefore, the Brans-Dicke type of extension indeed gives a more general theory of teleparallel gravity that includes both $F(T)$ gravity and teleparallel dark energy.

\section{General Analysis}\label{4}

We now analyze the more general action $S_{BD}$ of teleparallel gravity, together with a matter action $S_m$ that represents the gravitational source\footnote{The existence of the gravitational source does not change the discussion in
Sec.~\ref{2}.}, which is, unlike $S_{BD}$, assumed to be invariant under local Lorentz
transformation.
The variation of the action with respect to the tetrad $e^A{}_\nu$ gives one of the EoMs that, after contracted with $\kappa e_A{}^\mu / (2 |e|)$, reads
\begin{eqnarray}
&&{\cal{G}}^{\mu\nu}:=\( \partial_\lambda f \) S^{\mu\lambda\nu}
+f\frac{1}{e} \partial_\lambda \(e S^{\mu\lambda\nu} \)
-f T_{\lambda\rho}{}^\mu S^{\lambda\nu\rho}
-\frac{\alpha}{2} \( \partial^\mu \phi \)\(\partial^\nu \phi\) \nonumber\\
&&\qquad\qquad\qquad\qquad
- \frac{1}{4} g^{\mu\nu} \[ f(\phi)T-\alpha \( \partial_\lambda \phi \)\(\partial^\lambda \phi\)-V(\phi)\]-\frac{\kappa}{2}{\cal T}^{\mu\nu}=0,
\label{Eineq}
\end{eqnarray}
where the matter contribution is
\begin{eqnarray}
|e| {\cal T}^{\mu\nu} := -  e^{A\mu}\frac{\delta S_m}{\delta e^A{}_\nu}.
\end{eqnarray}
The other EoM is given by the variation with respect to $\phi$, which reads
\begin{eqnarray}
\Phi:=2\alpha \overset{\scriptscriptstyle L\;\; }{\nabla^\mu}\overset{\scriptscriptstyle L\;\; }{\nabla_\mu} \phi+f'(\phi)T-V'(\phi)=0.
\label{Eqphi}
\end{eqnarray}

To separate the extra DoFs from the ordinary ones associated with the metric, we decompose the 16 DoFs of a general tetrad $e^A{}_\mu$ into 6 DoFs of local Lorentz rotation $\Lambda^A{}_B$ and the remaining 10 in a more restricted tetrad $\ee^A{}_\mu$:
\begin{eqnarray}\label{L1}
 e^A{}_\mu = \Lambda^A{}_B \ee^B{}_\mu \, .
\label{rot}
\end{eqnarray}
Here, 16 components of tetrad $\ee^A{}_\mu$ are fixed by any 6 constraints that leave 10 DoFs of metric in tetrad $\ee^A{}_\mu$. 
Symmetrization of tetrad $\ee^B{}_\mu$ is one of the choices, 
with the understanding that the indices $B$ and $\mu$ are actually associated with different kinds of spaces.

The metric is independent of local Lorentz rotation $\Lambda^A{}_B$ and is therefore fully determined by the restricted tetrad $\ee^A{}_\mu$, i.e.,
\begin{eqnarray}
g_{\mu\nu}=\eta_{AB}e^A{}_\mu e^B{}_\nu
=\eta_{AB}\ee^A{}_\mu \ee^B{}_\nu \, .
\end{eqnarray}
Thus, $\ee^A{}_\mu$ contains all information of the metric with the accordant number of DoFs.
Some of the DoFs are not physical, and the DoF number can be reduced via a gauge choice. We consider a spacetime region where $\phi$ has no minimum and its gradient is timelike\footnote{In almost all cosmological models in generalized teleparallel theories, these conditions are satisfied.} and take the uniform-$\phi$ gauge where $\phi = \phi(t)$. Under this gauge one can write $\ee^A{}_\mu$ in a simpler form, with a suitable choice of $\Lambda^A{}_B$, as follows:\footnote{In the uniform density gauge, both the metric and $\ee^A{}_\mu$ have seven DoFs. }
\begin{eqnarray}
\ee^B{}_\mu = \left(
  \begin{array}{cc}
    N   & 0   \\
     0  & \ee^I{}_i   \\
  \end{array}
\right),\label{edec}
\end{eqnarray}
where $N$ plays the role of the lapse function.

In the following we analyze the dynamics of the
extra DoFs that are contained in the local Lorentz
rotation $\Lambda^A{}_ B$. Their dynamics are determined by the
EoMs (\ref{Eineq}-\ref{Eqphi}), where $\Lambda^A{}_
B$ is involved through $T^\mu{}_{\nu\lambda}$ and
$S_{\mu\nu\lambda}$. To single out the extra DoFs, we decompose
$T^\mu{}_{\nu\lambda}$ and $S_{\mu\nu\lambda}$ by separating
the parts which depend on $\Lambda^A{}_B$ from those which do not:
\begin{eqnarray}
&&T^\mu{}_{\nu\lambda}= \tt^\mu{}_{\nu\lambda} + \Delta T^\mu{}_{\nu\lambda},\\
&&S_{\mu\nu\lambda}=\SS_{\mu\nu\lambda}+\Delta S_{\mu\nu\lambda},
\end{eqnarray}
where the $\Lambda$-dependent parts read explicitly
\begin{eqnarray}
\Delta T^\mu{}_{\nu\lambda} &:=&
\ee_B{}^\mu \ee^C{}_\lambda \Lambda_A{}^B\partial_\nu \Lambda^A{}_C
-\ee_B{}^\mu \ee^C{}_\nu \Lambda_A{}^B\partial_\lambda \Lambda^A{}_C,\\
\Delta S_{\mu\nu\lambda} &:=&
\frac{1}{4}\Lambda_A{}^B\partial_\mu \Lambda^A{}_C
\(\ee_{B\nu}\ee^C{}_\lambda-\ee_{B\lambda}\ee^C{}_\nu\)
-\frac{1}{2}\Lambda_A{}^B\partial_\rho \Lambda^A{}_C
\(g_{\mu\nu}\ee_{B}{}^\rho \ee^C{}_\lambda- g_{\mu\lambda}\ee_{B}{}^\rho \ee^C{}_\nu\),
\end{eqnarray}
while the $\Lambda$-independent parts $\tt^\mu{}_{\nu\lambda}$
and $\SS_{\mu\nu\lambda}$ are constructed from $\ee^A{}_\mu$ under Eqs.\ (\ref{Tdef}) and (\ref{Sdef}),
respectively. The explicit components of
$\tt^\mu{}_{\nu\lambda}$ and $\SS_{\mu\nu\lambda}$ are listed
in Appendix \ref{A1}.

Using the relation between the Weitzenb\"{o}ck connection and the Levi-Civita connection, we can rewrite Eq.~(\ref{Eineq})  as
\begin{equation}
\( \partial_\rho f \) S^{\mu\rho\nu}
+\frac{1}{2} f G^{\mu\nu} -\frac{\alpha}{2} \( \partial^\mu \phi \)\(\partial^\nu \phi\)
+ \frac{1}{4} g^{\mu\nu} \[ \alpha \( \partial_\rho \phi \)\(\partial^\rho \phi\)+V\]
-\frac{\kappa}{2}{\cal T}^{\mu\nu}
 =0, \label{Eineq2}
\end{equation}
where $G^{\mu\nu}$ is the Einstein tensor based on the Levi-Civita connection. 
(For the detailed relation between the Weitzenb\"{o}ck connection and the Levi-Civita connection, see Sec.4 of \cite{Hayashi}.)
Only the first term on the left-hand side depends on the local Lorentz rotation $\Lambda^A{}_B$ while the other terms are invariant under the local Lorentz transformation.
Consequently, the extra DoFs appear in the EoM
(\ref{Eineq}) only through $\delta S_{\mu\nu\lambda}$, the
components of which read
\begin{eqnarray}
&& \Delta S_{tti} = \frac{N^2}{2}\Lambda_A{}^I\(\partial_k\Lambda^A{}_J\)\ee^J{}_i\ee_I{}^k,
\label{dStti}\\
&& \Delta S_{tij} = \frac{1}{4}\Lambda_A{}^I\(\partial_t\Lambda^A{}_J\)
                    \(\ee_{Ii}\ee^J{}_j-\ee_{Ij}\ee^J{}_i\),
\label{dStij}\\
&& \Delta S_{itj} = -\frac{N}{2}\Lambda_A{}^0\(\partial_i\Lambda^A{}_I\)\ee^I{}_j
                  +\frac{N}{2}\Lambda_A{}^I\(\partial_k\Lambda^A{}_0\) g_{ij}\ee_I{}^k,
\label{dSitj}\\
&& \Delta S_{ijk} = -\frac{1}{2N}\Lambda_A{}^0\(\partial_t\Lambda^A{}_I\)
                      \(g_{ij}\ee^I{}_k-g_{ik}\ee^I{}_j\)
  +\frac{1}{4} \Lambda_A{}^I\(\partial_i\Lambda^A{}_J\)\(\ee_{Ij}\ee^J{}_k-\ee_{Ik}\ee^J{}_j\)
  \nonumber\\ && \qquad\qquad\qquad\qquad
     -\frac{1}{2}\Lambda_A{}^I\(\partial_l\Lambda^A{}_J\)
         \(g_{ij}\ee_I{}^l\ee^J{}_k-g_{ik}\ee_I{}^l\ee^J{}_j\).  \label{dSijk}
\end{eqnarray}

Under the uniform-$\phi$ gauge, $f(\phi)$ depends only on time.
Therefore the first term in the EoM (\ref{Eineq2}) becomes $\(
\partial_t f \) S^{\mu t\nu}$, and only the terms in Eqs.\ (\ref{dStti}) and
(\ref{dSitj}) need to be considered. Note that $\delta S_{tti}$
and $\delta S_{itj}$ do not involve the time derivative of
$\Lambda^A{}_B$. As a result, the constant-$\phi$ hypersurface
is always the characteristic hypersurface for $\Lambda^A{}_B$, implying that it is a Cauchy horizon for the extra DoFs.
That is, regarding the variation of the extra DoFs in spacetime, the variation across the constant-$\phi$ hypersurface (i.e.\ the time evolution) cannot be uniquely determined by the EoMs with initial conditions, while the variation along the directions on the characteristic hypersurface can be determined by the EoMs with boundary conditions. Such variation on the hypersurface can be interpreted as a spacelike (therefore superluminal) propagation of the extra DoFs, a pathological type of behavior.
(We refer the readers to Appendix \ref{A2} for a quick introduction to the method of characteristics. For further reading, see e.g.~\cite{CH,Izumi:2013poa,Deser:2013eua} and more recently \cite{HR}.)

In TEGR, where $f(\phi)$ is constant, the first term on the left-hand side of Eq.~(\ref{Eineq2}) is algebraically zero. The equation then becomes local Lorentz invariant as expected. In other words, the DoFs associated with the breaking of local Lorentz invariance become gauge modes, and thus, the pathology in the extended theories which arises from these extra DoFs does not arise in TEGR. 
In contrast, $F(T)$ gravity with nonconstant $dF/dT$ is plagued by the same pathology as that in the Brans-Dicke type of extension. 
This can be seen by setting $\alpha=0$ in Eq.~(\ref{Eineq2}) since $F(T)$ gravity is a special case with $\alpha=0$, $f(\phi)=dF/d\phi$ and $V(\phi)=\phi (dF/d\phi)-F$, as stated below Eq.~(\ref{14}). In this case the first term on the left-hand side of Eq.~(\ref{Eineq2}), i.e.\ the carrier of the pathology, remains.

We emphasize the difference between the lack of kinetic terms in a constraint equation and the disappearance of a kinetic term on a characteristic hypersurface. 
The equations which, on all spacelike hypersurfaces, do not have a kinetic term are constraint equations, e.g., the Hamiltonian and momentum constraint equations in GR. 
The constraint equations render DoFs of propagation to be reduced. 
On the other hand, a hypersurface $\Sigma$ is characteristic if a kinetic term disappears only on $\Sigma$. 
That is, on generic hypersurfaces, the kinetic term shows up. 
Here, only when we choose the coordinate satisfying $\phi=\phi(t)$ to fix the time-constant hypersurface do the kinetic terms of  $\Lambda^A{}_B$ disappear. 
It exactly means that the time-constant hypersurface is characteristic. 
On the characteristic, the modes related to the disappearance of the kinetic terms must propagate 
(for the detailed discussion, see \cite{CH}).

\section{NonUniqueness of Time Evolution: a Concrete Example}\label{5}

In the previous section we have shown the existence of the
characteristic hypersurface, i.e.\ the constant-$\phi$
hypersurface, for local Lorentz rotation $\Lambda^A{}_B$ that carries the extra DoFs. The time evolution of the extra DoFs is therefore nonunique, an illness of the theory.
Nevertheless, in view of the discussion in Sec.~\ref{0}, if this illness does not affect physical
quantities such as the metric, one might argue that it is not physical and can be
ignored. In this section we  consider a simple case%
\footnote{A simpler case is the FLRW metric
(homogeneous and isotropic) that is included in the Bianchi type I metric. However, in this case the local Lorentz DoFs do not appear in the EoMs, and therefore, it is not clear how the pathological features could be physical.}
with the Bianchi type I metric and demonstrate such illness in the physical quantities, particularly in the gauge-invariant
metric components.

Recall that the Bianchi type I metric
\begin{eqnarray}
g_{\mu\nu}= \mbox{diag} \( -1 ,b^2, b^2, a^2 \)
\end{eqnarray}
describes a homogeneous but anisotropic spacetime.
It can be obtained from the tetrad with the ansatz: 
\begin{eqnarray}
\ee^0{}_t= 1, \qquad \ee^3{}_z=a(t), \qquad \ee^1{}_x=\ee^2{}_y=b(t),
\label{aee}
\end{eqnarray}
while all other components vanish.

For demonstration we consider a simple, special ansatz for
Lorentz rotation $\Lambda_A{}^B$:
\begin{eqnarray}
\Lambda_0{}^0=\Lambda_3{}^3= \cosh \( a(t) \theta (t) z\), \qquad
\Lambda_3{}^0=\Lambda_0{}^3= \sinh \( a(t) \theta (t) z\), \qquad
\Lambda_1{}^1=\Lambda_2{}^2= 1,
\label{aLam}
\end{eqnarray}
while all other components vanish. In EoMs the Lorentz DoFs always appear in the form of $\Lambda_A{}^B \partial_i
\Lambda^A{}_C$. Under the ansatz (\ref{aLam}), $\Lambda_A{}^B
\partial_i \Lambda^A{}_C$ can be nonzero only for
$(B,C,i)=(0,3,z)$ and $(3,0,z)$, both of which give $-a(t)\theta(t)$.
Therefore, the ansatz (\ref{aLam}) is consistent with the
homogeneous ansatz for the metric. The energy-momentum tensor
in this case has a diagonal form,
\begin{eqnarray}
{\cal T}_{\mu\nu}= \mbox{diag} \bigl( \rho(t), b^2 P(t), b^2 P(t), a^2 P_z(t) \bigr) ,
\end{eqnarray}
and satisfies the local conservation law,
$\overset{\scriptscriptstyle L\;\; } {\nabla^\mu} {\cal
T}_{\mu\nu}=0$.

The nontrivial components of EoMs in this case are as follows.
\begin{eqnarray}
{\cal G}_{tt}
&=& -\frac{1}{2} f H_b (2H_a +H_b) -\frac{\alpha}{4} (\partial_t \phi)^2-\frac{1}{4}V
-\frac{\kappa}{2} \rho =0,
\label{tt}\\
a^{-2}{\cal G}_{zz}
&=& f\partial_t H_b+\frac{3}{2}f H_b^2 + (\partial_t f)H_b
-\frac{\alpha}{4} (\partial_t \phi)^2+ \frac{1}{4}V- \frac{\kappa}{2}P_z=0, \\
{\cal G}_{pq}
&=& \frac{1}{2}g_{pq}\[
f\partial_t H_a+f \partial_t H_b + fH_a^2 +f H_a H_b +fH_b^2 +(\partial_t f)H_a
+(\partial_t f)H_b+\theta \partial_t f -\frac{\alpha}{2} (\partial_t \phi)^{2}
+\frac{1}{2}V-\kappa P\]=0,\nonumber\\
\\
\Phi &=& f'\(4H_aH_b+2H_b+4\theta H_b\)
+2\alpha\[\partial_t^2 \phi+ (H_a+2H_b) \partial_t \phi  \] -V'=0,
\label{Phi}
\end{eqnarray}
where
\begin{eqnarray}
H_a := \frac{\partial_t a}{a} \qquad \mbox{and} \qquad H_b := \frac{\partial_t b}{b} ,
\end{eqnarray}
i.e., the expansion rates in $z$ and
$(x,y)$ directions, respectively. Here seem to be four EoMs for four DoFs, including three ordinary DoFs, namely $a$, $b$ and $\phi$, and one extra DoF, $\theta$. However, Eqs.\ (\ref{tt})--(\ref{Phi}) are not independent because of the diffeomorphism invariance of the action. This invariance gives an identity:
\begin{eqnarray}
\overset{\scriptscriptstyle L\;\; }{\nabla^\mu}{\cal G}_{\mu\nu}
+ \frac{1}{4}\(\partial_\nu \phi\) \Phi=0.
\end{eqnarray}
With the ansatz (\ref{aee}) and (\ref{aLam}), only the $\nu=t$ component gives a nontrivial identity, i.e., a relation among
Eqs.\ (\ref{tt})--(\ref{Phi}). As a result, only three of the
EoMs are independent, one less than the DoFs.
Therefore, there is one DoF whose dynamics cannot be determined by EoMs.

In the following we concretely specify such uncontrolled DoF and present an illness in the solution. We introduce two physical variables:
\begin{eqnarray}
h=(H_a-H_b)/3 \qquad \mbox{and} \qquad H=(H_a+2H_b)/3,
\end{eqnarray}
characterizing the anisotropy and the average in expansion,
respectively. In a FLRW universe, $h$ is zero, and $H$
coincides with the Hubble expansion rate. In terms of $h$ and $H$, we write three independent EoMs as
\begin{eqnarray}
{\cal G}_{tt}&=&\frac{3}{2}(H-h)(H+h) -\frac{\alpha}{4} (\partial_t \phi)^2-\frac{1}{4}V
-\frac{\kappa}{2}\rho=0,
\label{eq1}\\
-2a^{-2}{\cal G}_{zz}+g^{pq}{\cal G}_{pq}&=&
3(\partial_t h) f + 9 H h f +3h(\partial_t \phi)f'+\theta (\partial_t \phi)f'
+\kappa\(P_z-P\)=0,
\label{eq2}\\
\Phi&=&f'\(6(H-h)(H+h)+4\theta(H-h)\)
+2\alpha\(\partial_t^2 \phi+ 3H \partial_t \phi  \) -V'=0.
\label{eq3}
\end{eqnarray}
Solving Eqs.\ (\ref{eq1}) and (\ref{eq2}) as the algebraic
equations of $H$ and $\theta$ and substituting the solution
into Eq.\ (\ref{eq3}), we obtain an equation for $\phi$ and $h$
with the sources $\rho$, $P_z$ and $P$ from matter. One can
treat this equation as a nonlinear
equation for $\phi$ and treat $h$ as an additional source and then solve this EoM of $\phi$ with any $h$. That is, $h$ is arbitrary and cannot be
determined by the EoMs even with a complete set of initial
conditions. For instance, one can construct a solution in which a
FLRW universe suddenly decays into an anisotropic universe or \emph{vice versa}.
Thus, the time evolution of the physical DoFs in the generalized theory (\ref{BD}) of teleparallel gravity is not unique. 
This remains true in $F(T)$ gravity, i.e., when $\alpha=0$.

In contrast, the pathology does not appear in TEGR, where $f'=0$. 
In this case the extra DoF, $\theta$, as always accompanied by the derivative of $f$, does not appear in the EoMs (\ref{eq1})--(\ref{eq3}). We accordingly have three well-behaved equations for three variables ($h$, $H$ and $\phi$), and therefore the time evolution is unique. Note that $\theta$ is related to the choice of tetrad, which is a gauge mode in TEGR.

\section{Summary and Discussion}\label{6}

We have investigated the (in)consistency of the Brans-Dicke type of extension of teleparallel gravity that includes both $F(T)$ gravity and teleparallel dark energy as special cases. We established the nonuniqueness of time evolution by finding the characteristic hypersurface for the extra DoFs that are additional to the ordinary DoFs associated with the metric. We further demonstrated a concrete example of such illness for a physical DoF in Bianchi type I spacetime.


In the general analysis, we attribute the extra DoFs to local Lorentz rotation so as to separate them from the ordinary metric DoFs and utilize the constant-$\phi$ gauge as well as the orthonomal coordinate (\ref{edec}). 
We then show the absence of the time derivative of local Lorentz rotation in the  the field equations. Note that, as we commented at the end of Sec.~\ref{4}, such an equation should not be mistaken for a constraint equation. Instead, it implies that the constant-$\phi$ hypersurface is the characteristic hypersurface on which the propagation of the extra DoFs can occur.
The speed of propagation can be infinite, a disaster from the viewpoint of causality.
The infinite speed of propagation renders the time evolution from the hypersurface nonunique, and therefore the theory has no predictability.


We have further demonstrated that the disastrous propagation can truly appear in physical quantities.
In Bianchi type I spacetime 
we find that the anisotropy in expansion can be an arbitrary function of time but not determined by the EoMs even with a complete set of initial conditions.
This implies that the time evolution is not unique from any constant-time hypersurface and
that there can be strange solutions where, for instance, a FLRW universe suddenly decays into an anisotropic universe or the other way around.
Since the anisotropy is physical, this disaster from the infinite-speed propagation is truly physical.

The undetermined function can be fixed by introducing boundary conditions on a timelike hypersurface.  This procedure is familiar in the context of an anti-de Sitter space, which is not globally hyperbolic and a boundary condition (usually a reflective one) needs to be imposed.\footnote{The problem for teleparallel theories investigated here is more serious in the sense that in an anti-de Sitter space one still has the well-posed Cauchy problem at least locally, but in our case, as already explained in \cite{Ong:2013qja}, even locally the time evolution into an \emph{infinitesimal} future cannot be uniquely determined.}
Nevertheless, the imposition of boundary conditions does not completely save the theory from the disastrous failure of lacking predictability.
The necessity of both the initial conditions on the spacelike hypersurface and the boundary conditions on the timelike hypersurface means that the information on the future boundary is required in order to solve the dynamics.
As in the case of an anti-de Sitter space, one may understand this as the theory being defined with the boundary condition at spatial infinity. However, even if so, in order to figure out the development of the local universe within the Hubble horizon, one must solve all of the dynamics in the entire spatial slice, including those outside the horizon.  It is certainly not desirable that one has to know what is happening at infinity in order to just calculate the orbital motion of an asteroid in the Solar System, for example.

Another problem is the existence of infinite-speed propagation.
These modes can propagate in any direction on the constant-$\phi$ hypersurface.
Therefore, any curve, including closed curves, on the constant-$\phi$ hypersurface can be causal.  In other words, closed causal curves abound in the theory and can be constructed much easier than in GR, much like they do in nonlinear massive gravity \cite{Deser:2013eua, Chien-I}.
Moreover, the closed causal curves can be infinitesimal, i.e.\ local objects, which are much more disastrous than the closed global timelike curves appearing in GR solutions such as the G\"odel solution.
As a result, the theory admits local acausality.
Note that acausality is a stronger statement than superluminality.
The latter simply means that the speed of light is not the upper bound for information propagation, which by itself needs not be problematic \cite{Bruneton,A,Geroch1}. By contrast, acausality is indeed unacceptable.

A problem may appear in the quantization of the theory where the time derivatives of the extra DoFs are absent.
The vanishing of the kinetic term always results in a strong coupling problem~\cite{ArkaniHamed:2002sp}.
With strong coupling the perturbative method in quantization cannot work, and the theory is out of control.
In fact, the validity of the \emph{classical} solutions is also in trouble.
Classical solutions are valid in the regime of the effective theory below an energy cutoff that, however, goes to zero when the kinetic term vanishes. Thus, the classical dynamics described by the action with vanishing kinetic terms is no longer reliable.

We now comment on the validity of the assumption that matter fields obey local Lorentz symmetry.
For a matter action which is not invariant under local Lorentz rotation $\Lambda_A{}^B$ and can accordingly involve the derivatives of $\Lambda_A{}^B$, the coupling between matter and the derivatives of $\Lambda_A{}^B$ should be very weak, as required by the stringent bounds from the experimental tests of the local Lorentz symmetry.
This extremely weak coupling, even if it exists, might simply change the position of the characteristic hypersurface a little bit but not drastically, and the existence of closed causal curves and other problematic features should still persist.

All of the disasters we discussed seem to stem from the extra DoFs which are activated by the breaking of the local Lorentz symmetry.\footnote{This of course does not mean that preserving local Lorentz symmetry will necessarily lead to the absence of extra DoFs.}
If the extra DoFs can be removed or excited in gentler ways so as to cause no problem, the theory may still be rescuable.
It might therefore be desirable to generalize teleparallel gravity in a way that does not involve extra DoFs. This gives a guideline for constructing a healthy generalized theory of teleparallel gravity.

\acknowledgments
The authors would like to thank James Nester for fruitful discussions and comments.
The authors are also grateful to Pisin Chen for much appreciated help and various supports.
Keisuke Izumi is supported by Taiwan National Science
Council under Project No.\ NSC101-2811-M-002-103,
and Je-An Gu under NSC101-2112-M-002-007.
Yen Chin Ong would like to thank Taiwan's Ministry of Education for 3-year-support via the Taiwan Scholarship.

\appendix
\section{Some Tensor Components}\label{A1}
The components of $\tt^\mu{}_{\nu\lambda}$,
$\SS_{\mu\nu\lambda}$ and other relevant quantities (under the
uniform-$\phi$ gauge) are as follows.
\begin{eqnarray}
&&\tt^t{}_{ti}=-\partial_i \( \log N\),\\
&&\tt^t{}_{ij}=0,\\
&&\tt^i{}_{tj}= \ee_I{}^i \partial_t \ee^I{}_j, \\
&&\tt^i{}_{jk}=\bar T^i{}_{jk}.\\
&&\tt_t:=\tt^\mu{}_{t\mu}= \ee_I{}^i \partial_t \ee^I{}_j, \\
&&\tt_i:=\tt^\mu{}_{i\mu}=\partial_i \( \log N\)+\bar T_i,\\
&&\SS_{tti}=-\frac{N^2}{2}\bar T_i,\\
&&\SS_{tij}=\frac{1}{4}\( \ee_I{}_i \partial_t \ee^I{}_j -\ee_I{}_j \partial_t \ee^I{}_i\),\\
&&\SS_{itj}=\frac{1}{4}\(\ee_I{}_i \partial_t \ee^I{}_j +\ee_I{}_j \partial_t \ee^I{}_i \)
             -\frac{1}{2}g_{ij} \ee_I{}^k \partial_t \ee^I{}_k,\\
&&\SS_{ijk}=\bar S_{ijk}+\frac{1}{2}g_{ij}\(\partial_k \log N\) -\frac{1}{2}g_{ik}\(\partial_j \log N\),
\end{eqnarray}
where
\begin{eqnarray}
&&\bar T^i{}_{jk} :=\ee_I{}^i \partial_j \ee^I{}_k -\ee_I{}^i \partial_k \ee^I{}_j,\\
&&\bar T_i := \bar T^j{}_{ij},\\
&&\bar S_{ijk}:=-\frac{1}{4}\bar T_{jki}+\frac{1}{4}\bar T_{kji}+\frac{1}{4}\bar T_{ijk}
  +\frac{1}{2}g_{ij} \bar T_k-\frac{1}{2}g_{ik}\bar T_j.
\end{eqnarray}


\section{Method of Characteristics}\label{A2}

Consider a quasilinear $N$th-order differential equation representing the EoM for a physical system:
\begin{equation}
a_N^{{\mu_1}{\mu_2}\cdots {\mu_N}}\partial_{\mu_1}\partial_{\mu_2} \cdots \partial_{\mu_N} \phi + F(\phi,\partial_\mu \phi, \cdots,\partial_{\mu_1} \partial_{\mu_2} \cdots \partial_{\mu_{N-1}} \phi)= 0,\label{PDE}
\end{equation}
where the first term is linear with the $N$th-order derivative of $\phi$, and $a_N^{{\mu_1}{\mu_2}\cdots {\mu_N}}$ and $F$ are functions of $\phi$ and its derivatives up to the $(N-1)$th-order.
This equation can be decomposed as
\begin{equation}
a^{tt\cdots t}_N \partial_t^N \phi + f\left(\partial_t^{N-1}\phi, \partial_t^{N-2}\phi, \cdots, \partial_t^{N-1}\partial_i\phi, \cdots, \phi \right) = 0,
\end{equation}
where the $N$th-order time derivative is singled out and put in the first term, while the others are represented by the function $f$.

Consider the time evolution from an initial spacelike hypersurface at some time $t=t_0$. That is, we start with values of the fields $\left\{\phi, \partial_t \phi, \cdots, \partial_t^{N-1}\phi \right\}$ at $t=t_0$ and use the EoM to determine $(\partial_t^N \phi) (t_0)$, so that we can in turn determine $(\partial_t^{n-1}\phi)(t_0 + \Delta t)$ for $n = N, N-1,\cdots,1$ in the infinitesimal future via
\begin{equation}
(\partial_t^{n-1}\phi)(t_0 + \Delta t) = (\partial_t^{n-1}\phi) (t_0) + (\partial_t^n \phi)(t_0) \cdot \Delta t \, .
\end{equation}
If $a^{tt \cdots t}=0$ at $t_0$, the EoM becomes singular and $\partial_t^N \phi (t_0)$ can have any value. Time evolution thus becomes nonunique. 

More generally, consider a spacelike hypersurface $\Sigma$ with a timelike normal vector $\xi^\mu$. If $\xi^\mu$ satisfies
\begin{equation}
\label{PS}
a_N^{\mu_1 \mu_2 \cdots \mu_N} \xi_{\mu_1} \xi_{\mu_2} \cdots \xi_{\mu_N} = 0,
\end{equation}
the hypersurface $\Sigma$ is called the \emph{characteristic hypersurface} and $\xi^\mu$ the \emph{characteristic direction}.
Time evolution beyond such hypersurface is not uniquely determined by initial conditions, i.e.\ $\Sigma$ is the edge of Cauchy development. Accordingly, $\phi$ can be discontinuous across $\Sigma$, and physically one can visualize $\Sigma$ as a propagating shock wave front, which
is a three-dimensional hypersurface tangent to the two-dimensional shock wave front and to the direction of its propagation.
In the theory of differential equations, the left-hand side of Eq.~(\ref{PS}) is called the \emph{principal symbols} for the system (\ref{PDE}). By studying the solutions of Eq.~(\ref{PS}), i.e.\ the zeros of the principal symbols, one can find out how the shock wave front propagates. In particular, a timelike characteristic direction, like the case we have in the present paper, means that the propagation on $\Sigma$ is superluminal. A physically acceptable shock wave front should be timelike and thus has a spacelike characteristic direction. In contrast, a spacelike (superluminal) shock is odd and problematic.

We emphasize that the characteristic method is a standard analysis in differential equations. It is very powerful in revealing problematic (superluminal and acausal) propagation  without explicitly finding a solution which exhibits acausality, such as the G\"odel solution in GR famous for its closed timelike curve.


\end{document}